\documentstyle[12pt]{article}

\input psfig

\def\reference{\parskip 0pt\par\noindent\hangindent 0.5 truecm}

\begin{document}

\title{The Spectra of Red Quasars}

\author{Paul J. Francis $^{1,2}$ \and
Catherine L. Drake $^1$ \and 
Matthew T. Whiting $^{3}$ \and
Michael J. Drinkwater $^{3}$ \and
Rachel L. Webster $^{3}$
}

\date{}
\maketitle

{\center
$^1$ Research School of Astronomy and Astrophysics, Australian National
University, Canberra ACT 0200\\pfrancis,cdrake@mso.anu.edu.au\\[3mm]
$^2$ Joint appointment with the Department of Physics, Faculty of 
Science, Australian National University\\[3mm]
$^3$ School of Physics, University of Melbourne, Victoria 3010
\\mwhiting, m.drinkwater, rwebster@physics.unimelb.edu.au\\
[3mm]
}

%
\begin{abstract}

We measure the spectral properties of a representative sub-sample of 187
quasars, drawn from the Parkes Half-Jansky, Flat-radio-spectrum Sample
(PHFS). Quasars with a wide range of rest-frame optical/UV continuum slopes
are included in the analysis: their colours
range from $2 < B-K < 7$. We present composite spectra of red and blue
subsamples of the PHFS quasars, and tabulate their emission-line properties.

The median H$\beta$ and [O~III] emission-line equivalent widths of the red 
quasar sub-sample  
are a factor of ten weaker than those of the blue quasar sub-sample. 
No significant
differences are seen between the equivalent width distributions of the 
C~IV, C~III] and Mg~II lines. Both the colours and the emission-line
equivalent widths of the red quasars can be explained by 
the addition of a featureless red synchrotron continuum component to an 
otherwise normal blue quasar spectrum. The red synchrotron component must have
a spectrum at least as red as a power-law of the form 
$F_{\nu} \propto \nu^{-2.8}$. The relative strengths of the blue and red 
components span two orders of magnitude at rest-frame 500nm. The blue 
component is weaker relative to the red component in low optical luminosity 
sources. This suggests that the fraction of accretion energy going into 
optical emission from the jet is greater in low luminosity quasars. This
correlation between colour and luminosity may be of use in cosmological 
distance scale work.

This synchrotron model does not, however, fit $\sim 10$\% of the quasars, 
which have both red colours and high equivalent width emission-lines. We 
hypothesise that these red, strong-lined quasars have intrinsically weak 
Big Blue Bumps.

There is no discontinuity in spectral properties between the BL Lac objects
in our sample and the other quasars. BL Lac objects appear to be the red,
low equivalent width tail of continuous distribution. The synchrotron
emission component only dominates the spectrum at longer wavelengths, so
existing BL Lac surveys will be biassed against high redshift objects.
This will affect measurements of BL Lac evolution.

The blue PHFS quasars have significantly higher equivalent width C~IV, 
H$\beta$ and [O~III] emission than a matched sample of optically selected QSOs.

\end{abstract}

{\bf Keywords:}
Quasars: general --- Quasars: Emission Lines --- BL Lacertae Objects:
General
\bigskip

\section{INTRODUCTION}

The rest-frame optical/UV emission of radio quiet QSOs is dominated
by strong broad emission lines and by extremely blue continuum emission
(the Big Blue Bump, which peaks in the far-UV, eg. Malkan \& Sargent 
1982). The emission of flat radio 
spectrum radio-loud quasars 
is often quite different. It is now well established that many have much
redder colours than traditional optically selected QSOs (eg. Rieke,
Lebofsky \& Wisniewski 1982, Webster et al. 1995). A few also have extremely 
low equivalent width emission lines, and are known as BL Lac objects.

What is responsible for these differences? In this paper, we address this 
question, using the spectra of a representative sub-set of flat radio 
spectrum sources from the Parkes Half-Jansky Flat-Spectrum survey (PHFS). 
This survey 
consists of 323 sources with flux densities at 2.7 GHz of greater than 0.5 
Jy, and radio spectral indices $\alpha$ ($F_{\nu} \propto \nu^{\alpha}$) 
with $\alpha > -0.5$ as measured between 2.7 and 5.0 GHz (Drinkwater et 
al. 1997). The PHFS quasars lie at redshift $0 < z < 4$.

Francis, Whiting and Webster (2000, hereafter FWW) obtained quasi-simultaneous
optical and near-IR photometry for a random subset of 157 PHFS sources. They 
showed that 
the spectral energy distributions of red quasars are diverse. They divided the
sources, on the basis of their colours, into three classes.

\begin{enumerate}

\item Dusty quasars. About 10\% of the sample fall into this category.
They have the strongly curved spectral energy distributions expected
from dust obscuration, and show reddened line ratios.
Their spectra are discussed by FWW.

\item Galaxies. Another 10\% of the sources are spatially resolved, and
have spectra typical of elliptical galaxies. These sources are discussed
by Masci, Webster and Francis (1997).

\item Pseudo power-law sources. About 80\% of the sources have approximately
power-law spectral energy distributions, extending from the rest-frame UV to 
the near IR. These power-law continua vary in slope
between $F_{\nu} \propto \nu^{\sim 0}$ and $F_{\nu} \propto \nu^{\sim -2}$.
These are the sources we discuss in this paper.

\end{enumerate}

What are these pseudo power-law sources, and what is responsible for their
diversity of continuum slopes? Three models have been proposed. Firstly,
red pseudo-power-law sources might have intrinsically weaker Big Blue Bumps
(eg. McDowell et al. 1989). Secondly, all quasars might have similar Big Blue
Bumps, but in the redder quasars this emission might be partially absorbed by
dust (eg. Webster et al. 1995). The dust properties would have to be unusual
not to induce curvature in the spectra. Thirdly, all quasars might have 
similar Big Blue Bumps, but in the redder quasars, this Big Blue Bump emission
is swamped by a red synchrotron emission component from the radio jet
(eg. Serjeant \& Rawlings 1997).

In this paper, we test these models against the {\em spectra} of the PHFS 
pseudo power-law sources.
In Section~\ref{sample} we discuss our spectra of the PHFS quasars, and
show that they represent a reasonably representative sub-sample. In
Section~\ref{analysis} we describe in detail our measurements of the spectra, and in
Section~\ref{results} we show the results. The predictions of the three
models for the emission-line properties are discussed in Section~\ref{models}.
We show, in Section~\ref{discussion},
that the spectra are mostly consistent with the synchrotron model, and
derive constraints on this model. Finally, conclusions are drawn in
Section~\ref{conclusions}.

\section{The Sample\label{sample}}

We obtained spectra for 77 of the PHFS sources with the Anglo-Australian 
Telescope (AAT) and the Siding Spring 2.3m telescope (Drinkwater et al.
1997). These were combined with 25 unpublished spectra drawn from the
AAT archives, and 86 spectra drawn from the compilation of Wilkes et al.
(1983). We thus have spectra of 187 of the 323 sources in the PHFS.

\begin{figure}
 \begin{center}
 \psfig{file=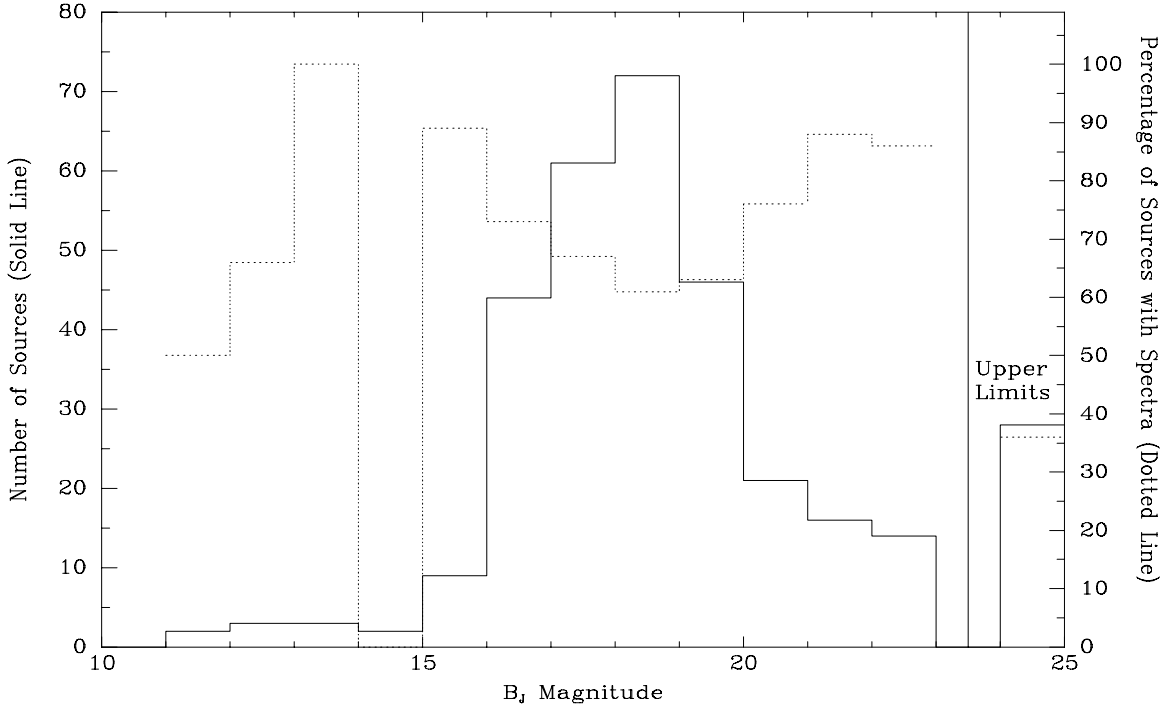,width=160mm}
 \caption{The number of PHFS sources as a function of $B_J$ magnitude
(solid line, left axis) and the percentage of PHFS sources for which we have 
digital spectra (dotted line, right axis). Sources for which we only have an 
upper limit
on their $B_J$ magnitude are shown in the right panel.\label{completeness}}
 \end{center}
\end{figure}

In Fig~\ref{completeness}, we show the number of PHFS sources for
which we have spectra as a function of $B_J$ magnitude. Note that the
fraction with spectra is reasonably uniform. 

The quality and wavelength coverage of the spectra is very diverse. The
signal to noise ratio of most quasars was too poor to allow measurements of
line widths or profiles. As most are not spectrophotometric, we did not
attempt to extract continuum parameters or line ratios from the spectra.
The quality and diversity of the spectra can best be appreciated by looking
at the sample spectra in Drinkwater et al. (1997).

\subsection{Sub-samples}

Spatially extended sources were excluded from this paper. Images from the 
UK Schmidt
and Palomar sky survey images (IIIa-J) were automatically classified, using 
the COSMOS and APM scanners. The automated
classification was checked against our own deeper imaging. We classified
18 of the PHFS sources with spectra as spatially extended:  they are
discussed by Masci et al. (1997).

The remaining point-like sources were analysed on the basis of their
optical/near-IR colours. The quasi-simultaneous broad-band magnitudes
of FWW were used. This excluded half of the quasars with spectra, as no 
photometry was available for them. Sources with $B-I > 2.6$ were excluded
from the analysis. These sources have the convex continuum spectral energy 
distributions expected from dust absorption, and their spectra were discussed
in FWW.

\begin{figure}
 \begin{center}
 \psfig{file=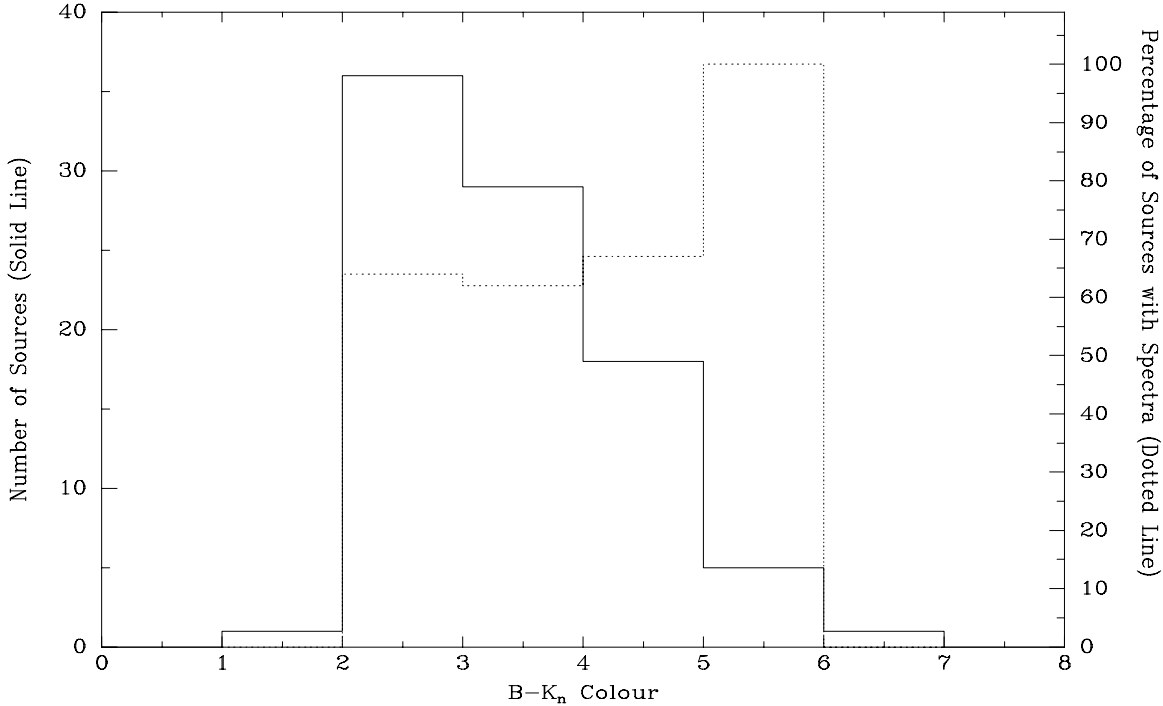,width=160mm}
 \caption{The number of spatially unresolved PHFS sources observed by FWW, with
$B-I<2.6$, as a function of their
$B-K_n$ colour (solid line, left axis). The percentage for which we have 
digital spectra is shown as the dotted line (right axis).
\label{completeness_bk}}
 \end{center}
\end{figure}

In Fig~\ref{completeness_bk} we show the
percentage of these remaining pseudo power-law sources for which we have 
spectra, as a function of their $B-K_n$ colour. Note once again that the 
subsample with spectra is reasonably representative.

We defined two subsamples of these pseudo-power-law sources. They were
defined on the basis of their rest-frame colours. We measured
these colours from our broad-band photometry, and not from the
(non-spectrophotometric) spectra. The flux densities at 340nm and 750nm
were calculated by extrapolation between the photometric data points from 
FWW.

\begin{itemize}

\item The Blue Sample. 22 sources with spectra. Flux density ($F_{\lambda}$) 
at rest-frame 340nm which exceeds that at 750nm by more than 0.45 dex 
(a factor of 2.8)

\item The Red Sample. 20 sources with spectra. Flux density ratio of less
than 0.45 dex between these wavelengths.

\end{itemize}

This dividing line corresponds to a power-law of index $F_{\nu} \propto
\nu^{-0.7}$ between these two wavelengths, and was chosen to bisect the
sample.

As a control sample, the Large Bright QSO Sample (LBQS) was used
(Morris et al. 1991 and refs. therein). This sample of 1053 optically
selected QSOs is well matched in both redshift and optical luminosity
to the PHFS. Broad absorption line QSOs were excluded from the comparison.

\section{Analysis\label{analysis}}

Error arrays were not available for most of the sample spectra. We
therefore estimated them as follows. For each pixel, we fit a straight
line to the fluxes in the 15 pixels centred on it. The sum of the squares
of the residuals from this fit was divided by 13 to estimate
the error variance in the pixel. We divided by 13 rather than 15 to allow
for the two degrees of freedom in the straight-line fit.

We iteratively fit a 4th order Chebyshev polynomial to the continuum
of each quasar. Wavelength regions with strong emission-lines were
excluded from the fitting. The fits were iterated three times, with
regions more than $2 \sigma$ above or $4 \sigma$ below each fit excluded
from the next. The results were checked by eye: these fits were subjectively
judged to be good for 95\% of the quasars. The remaining 5\% were fit 
interactively.

Emission-line equivalent widths were obtained by dividing the flux above
these continuum fits by the continuum level, integrated over the
wavelength regions in Table~\ref{linelimits}. Only the strongest
emission-lines were measured, as the signal-to-noise ratio was too poor
to allow measurement of the weaker lines. Ly$\alpha$ was not measured
due to the difficulty in defining the continuum under it, and the small
number of red quasars at high enough redshift. The [O~III]
equivalent widths are for the stronger component (the 500.7 nm line)
only. All the continuum fits
and line equivalent widths were checked by eye, and about 10\% of
the automated measurements were replaced with interactive ones.

\begin{table}
\begin{center}
\caption{Emission-line Integration Limits\label{linelimits}}
\begin{tabular}{lcc}
\hline 
Line & Lower Rest Wavelength Limit & Upper Rest Wavelength Limit \\
\hline \\
C~IV     & 149 nm & 159 nm \\
C~III]   & 183 nm & 196 nm \\
Mg~II    & 268 nm & 290 nm \\
H$\beta$ & 476 nm & 493 nm \\ 
$[$O~III$]$  & 498 nm & 505 nm \\ \hline \\
\end{tabular}
\end{center}
\end{table}

Composite quasar spectra were obtained as follows. Each individual spectrum 
was divided through by the continuum fit and shifted to its rest-frame,
using interactively defined redshifts. The spectra were then rebinned
and co-added. All spectra were equally weighted in the co-add, regardless
of their signal-to-noise ratio. This degrades the signal-to-noise ratio
of the final composite spectrum, but ensures that it is not biassed towards
the properties of the quasars with better quality spectra. The fluxes at 
very blue and very red wavelengths were given
a low weighting, to reflect their generally poor signal-to-noise ratios.
Only wavelength regions to which at least five individual quasars 
contributed were included in the composite spectra.

\subsection{Sources without Redshifts}

Three sources have no redshift in Drinkwater et al. (PKS 0048$-$097,
PKS 0829$+$046 and PKS1156$-$094). Falomo (1991) measured a
redshift of 0.18 for PKS 0829$+$046, which we adopt. Both the other
sources have red colours and would lie in the red sample regardless
of their redshift.

We estimated upper limits on the emission-line equivalent widths of
PKS 0048$-$097 and PKS 1156$-$094 as follows. The biggest wavelength gap
between strong emission lines is between Mg~II and H$\beta$. The ratio
of their wavelengths is 1.74. Thus any wavelength range with an end 
wavelength more than 1.74 times its starting wavelength must contain
at least one strong line. In both spectra, the least noisy wavelength 
range of this width was taken. The $3 \sigma$ limit for a line of width
$5000 {\rm km\ s}^{-1}$ was computed as a function of wavelength throughout
this region, and the highest value taken as our limit.

We place a $3 \sigma$ upper limit of 0.17nm for PKS 0048$-$097. The spectrum 
of PKS 1156$-$094
is poor, and we can only place a $3 \sigma$ upper limit of 6.6nm on the 
equivalent width of any emission line.

\subsection{Luminosities and Rest-frame Colours}

Continuum luminosities were calculated at particular rest-frame wavelengths,
by linear extrapolation between adjacent redshifted photometric bands from FWW.
We assume that $H_0 = 70 {\rm km\ s}^{-1}{\rm Mpc}$, $\Omega_{\rm matter} = 
0.3$ and $\Omega_{\Lambda} = 0.7$ throughout. Radio luminosities were
calculated in the same way, using radio flux densities at 1600, 2468, 
4800 and 8640 MHz obtained by Webster et al. (in preparation).
Emission-line luminosities were calculated by scaling the spectra to match
the broad-band photometry of FWW extrapolated to the line wavelength.

\section{Results\label{results}}

\subsection{Composite Spectra}

\begin{figure}

 \psfig{file=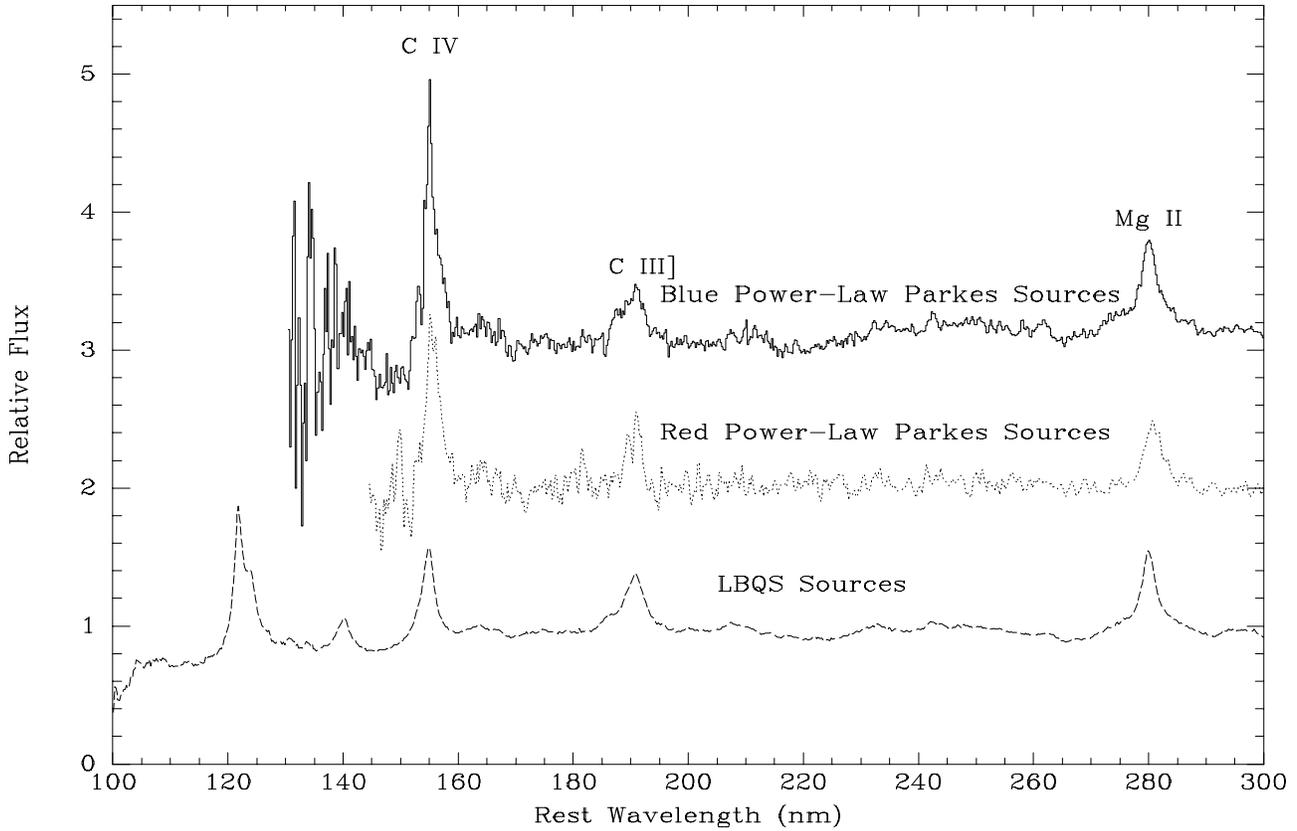,width=180mm}
 \caption{The blue end of the composite spectra of blue
PHFS sources, red PHFS sources, and optically selected sources
from the LBQS. The PHFS spectra are vertically offset by 1 (red) and 2 (blue)
for clarity. Note that the spectra were divided through by a continuum
fit before being co-added.\label{bluecomp}}

\end{figure}

\begin{figure}
 \psfig{file=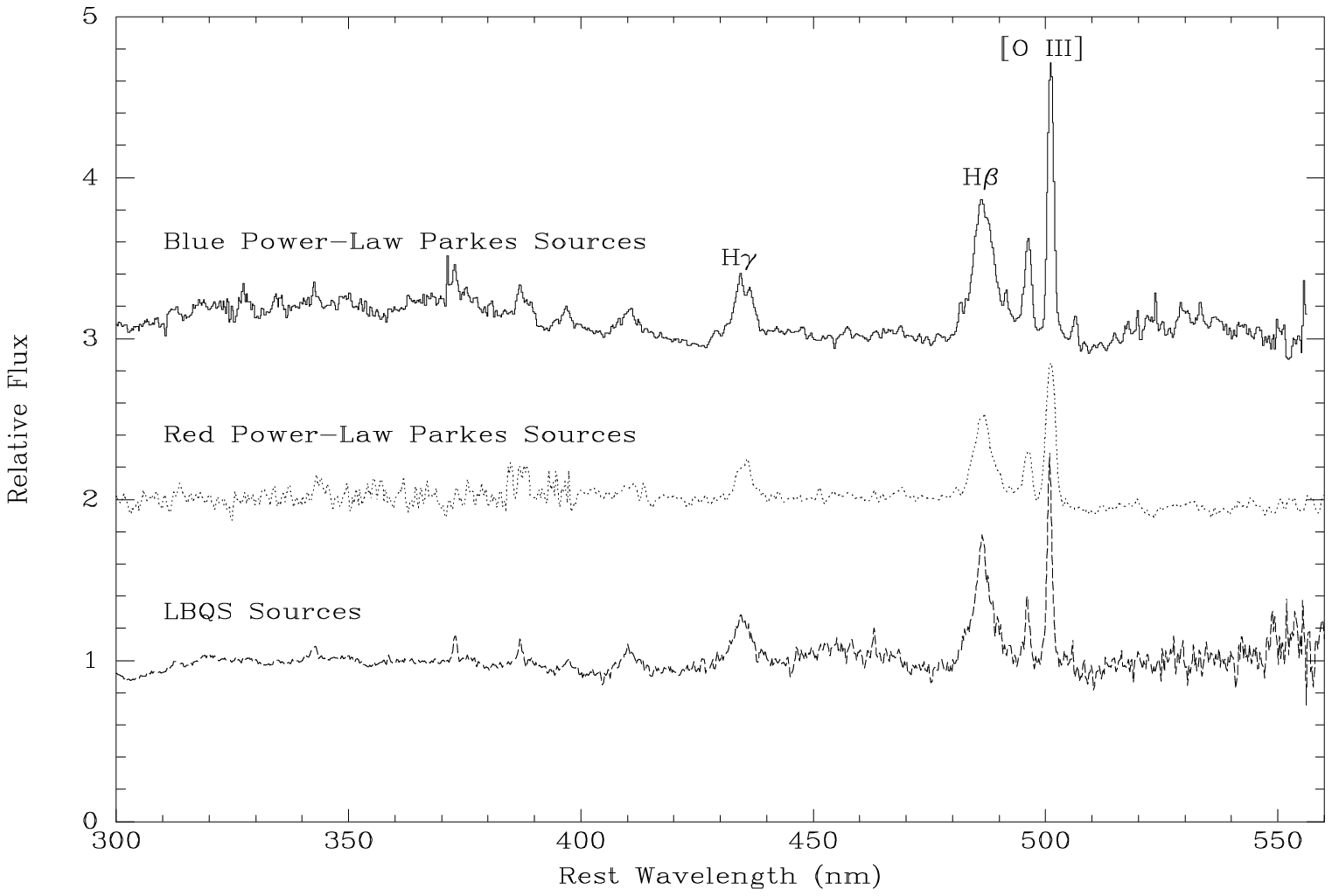,width=180mm}
 \caption{The red end of the composite spectra of blue
PHFS sources, red PHFS sources, and optically selected sources
from the LBQS. The PHFS spectra are vertically offset by 1 (red) and 2 (blue)
for clarity. Note that the spectra were divided through by a continuum
fit before being co-added.\label{redcomp}}

\end{figure}

The composite spectra of blue and red PHFS quasars (excluding the two 
without measured redshifts), and of the optically
selected LBQS QSOs, are shown in Figs~\ref{bluecomp} and \ref{redcomp}.
A number of differences can be seen. The red PHFS composite has much
weaker Mg~II, Balmer and [O~III] lines than the blue PHFS composite.
The C~III] and C~IV lines, however, are more similar in strength.
The LBQS composite more closely resembles the blue PHFS composite, but
its Ly$\alpha$, C~IV, H$\beta$ and [O~III] lines are weaker.
The reality of these differences will be addressed in the next section.

\subsection{Equivalent Width Measurements}

\begin{figure}
\psfig{file=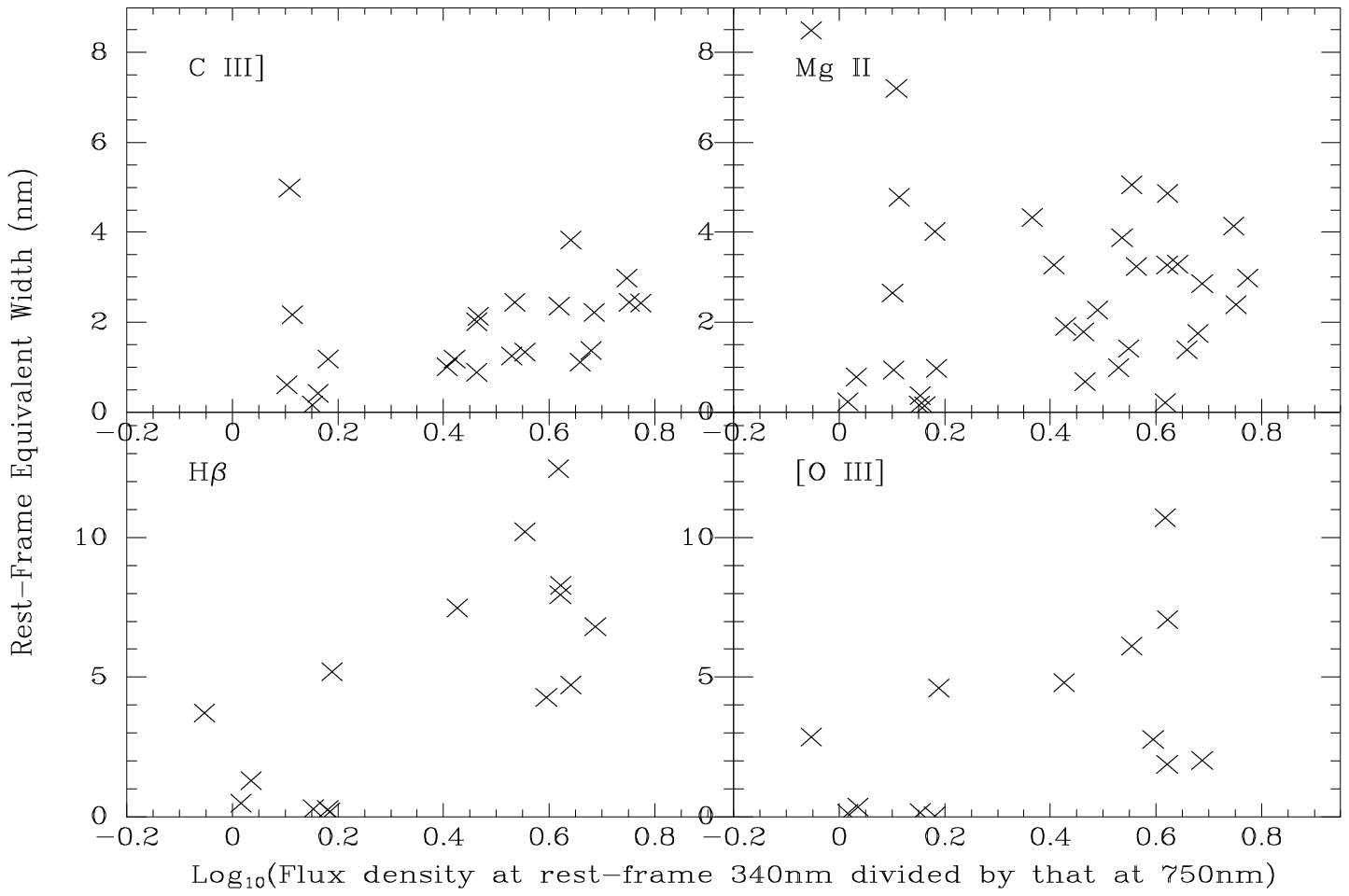,width=180mm}
\caption{The rest-frame equivalent widths of four strong emission
lines as a function of continuum slope. Continuum slope is measured
as the flux density ($F_{\lambda}$) at rest-frame 340nm divided by that at 
750nm.
\label{ewbk}}

\end{figure}

\begin{figure}
 \psfig{file=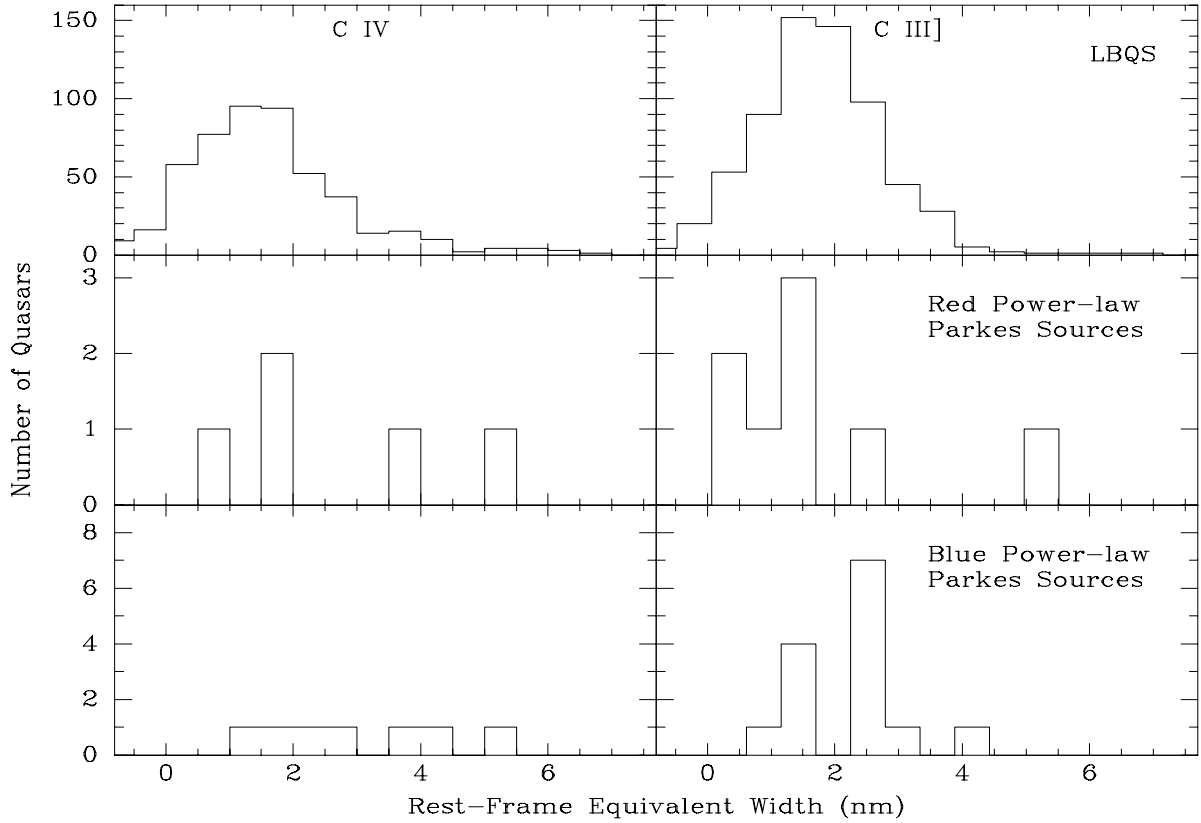,width=180mm}
 \caption{Rest-frame equivalent width histograms for C~IV
(left panels) and C~III] (right panels). Optically 
selected LBQS QSOs are shown in the top panels, red PHFS quasars
in the middle panels and blue PHFS QSOs in the bottom panels.\label{hist1}}

\end{figure}

\begin{figure}
 \psfig{file=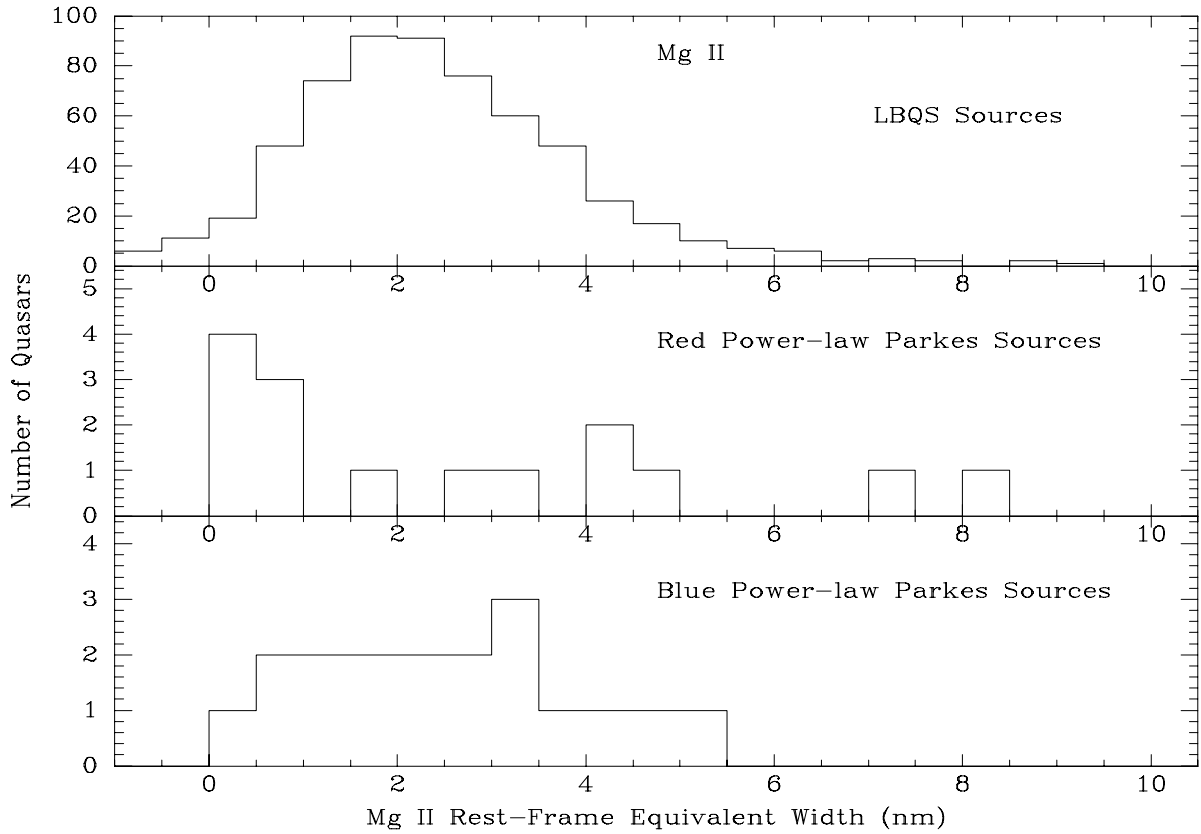,width=180mm}
 \caption{Rest-frame equivalent width histograms for Mg~II.
Optically selected LBQS QSOs are shown in the top panel, red PHFS quasars
in the middle panel and blue PHFS QSOs in the bottom panel.\label{mghist}}

\end{figure}

\begin{figure}
 \psfig{file=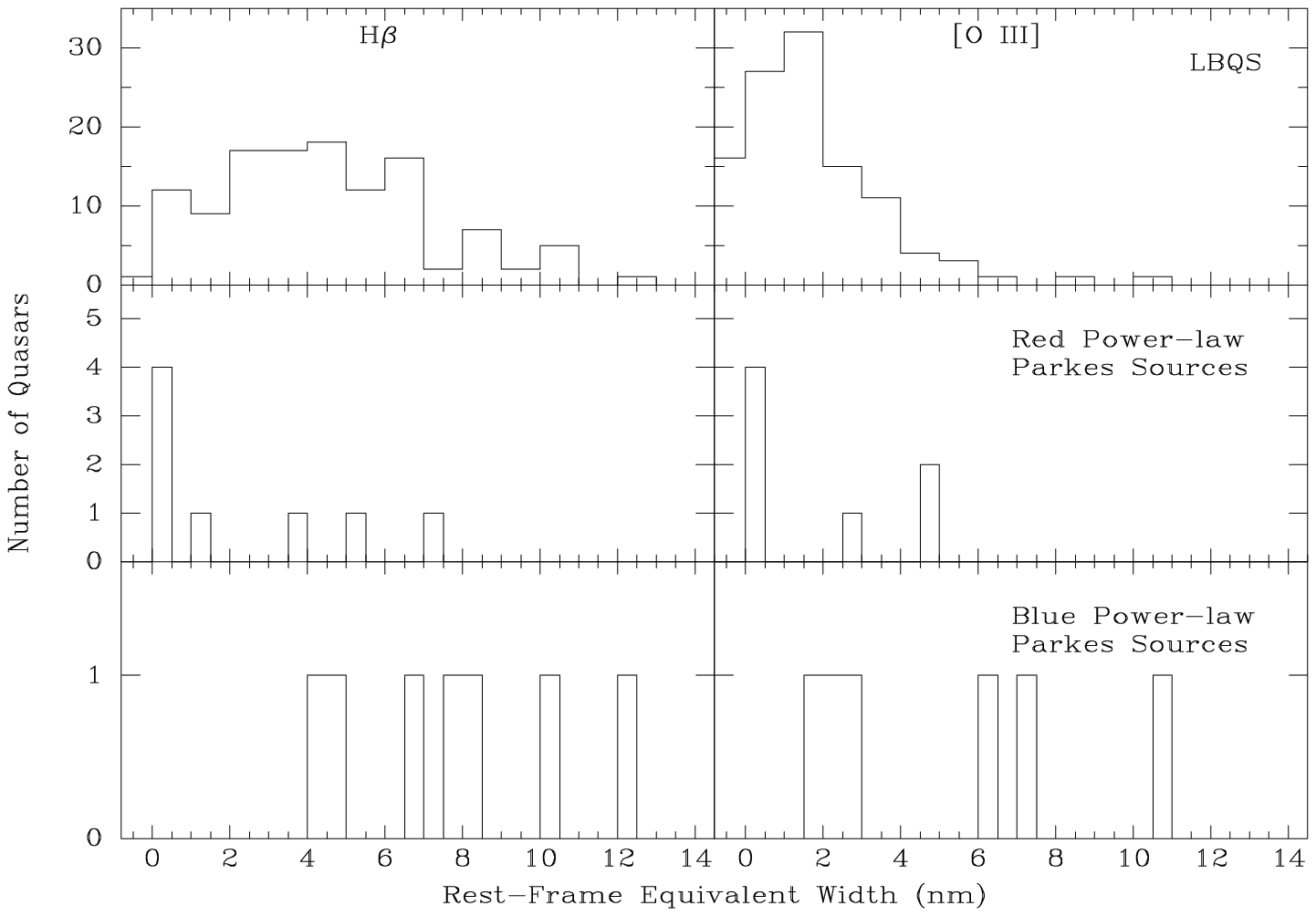,width=180mm}
 \caption{Rest-frame equivalent width histograms for H$\beta$
(left panels) and [O~III] (right panels). Optically 
selected LBQS QSOs are shown in the top panels, red PHFS quasars
in the middle panels and blue PHFS QSOs in the bottom panels. Note the
small number statistics in the lower panel.\label{hist2}}

\end{figure}

The equivalent width measurements can be found in a machine readable
table, included in the electronic edition of this paper. 
Equivalent widths are plotted against continuum slope in Fig~\ref{ewbk}.
Histograms of equivalent widths for the blue and red sub-samples, and
for the LBQS, are shown in Figs~\ref{hist1}, \ref{mghist} and 
\ref{hist2}.
The two PHFS quasars without measured redshifts (both in the
red sample) are not included.

Are the differences between the red and blue sub-samples real? The correlation 
between the equivalent width of 
H$\beta$ and the continuum slope is significant at the 99.2\% level, using
Spearman's Rank Correlation test. A Kolmogorov-Smirnov test confirms that
the H$\beta$ equivalent width distributions of the two sub-samples 
are different at the 97\% confidence level.
The C~III] equivalent width distributions are marginally inconsistent
(at the 95\% level). There are no other differences significant at the
95\% confidence level.

The two quasars without measured redshifts were excluded from this
analysis. If we assume that they are of low enough redshift that H$\beta$
would have been seen, they increase the significance of the difference.

\subsection{Line Ratios}

No correlation was found between line flux ratios and colour. The
correlation predicted by a dust model is, however, weak enough to be 
consistent with the observations. We have wide wavelength coverage,
spectrophotometric spectra for three of the reddest pseudo power-law sources,
however, from which we can measure Balmer decrements. In all three,
the ratio of H$\alpha$ to H$\beta$ is around 3.5, consistent with
no dust reddening.

\subsection{Correlations with Luminosity}

\begin{figure}
 \psfig{file=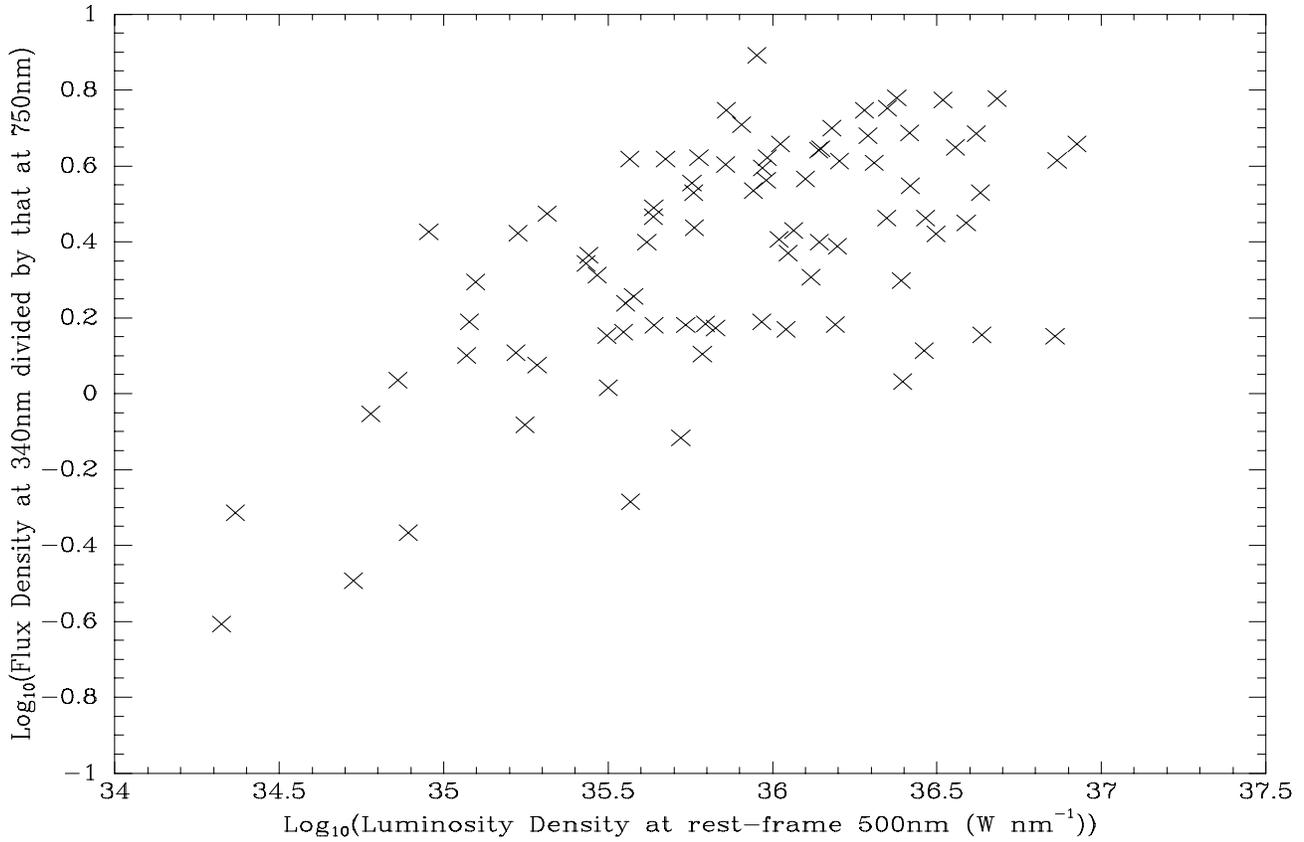,width=180mm}
 \caption{The continuum slope, measured as a function of the continuum 
luminosity at rest-frame 500nm. The continuum slope is measured by the
flux ($F_{\lambda}$) at rest-frame 340 nm divided by that at 750nm.
\label{slope_lum}}

\end{figure}

The continuum slope of the PHFS samples correlates with the
optical continuum luminosity (Fig~\ref{slope_lum}). Spearman's
Rank Correlation Test shows that this correlation is significant with   
99.999\% confidence. No significant
correlation is seen between continuum slope and {\em radio} luminosity.

As the PHFS is a flux limited sample, redshift and luminosity are
strongly correlated. Is this really a correlation against luminosity, or
could it be a correlation against redshift? To test this, the sample was
subdivided into redshift bins. The median luminosity of the sources in each bin
was calculated, and subtracted from all the sources in that bin. The resultant 
differential luminosities still
correlated with colour, with 95\% confidence. When the process was
repeated with luminosity bins, no significant residual correlation
was seen. We therefore conclude that the physical correlation is between
colour and luminosity.

\begin{figure}
 \psfig{file=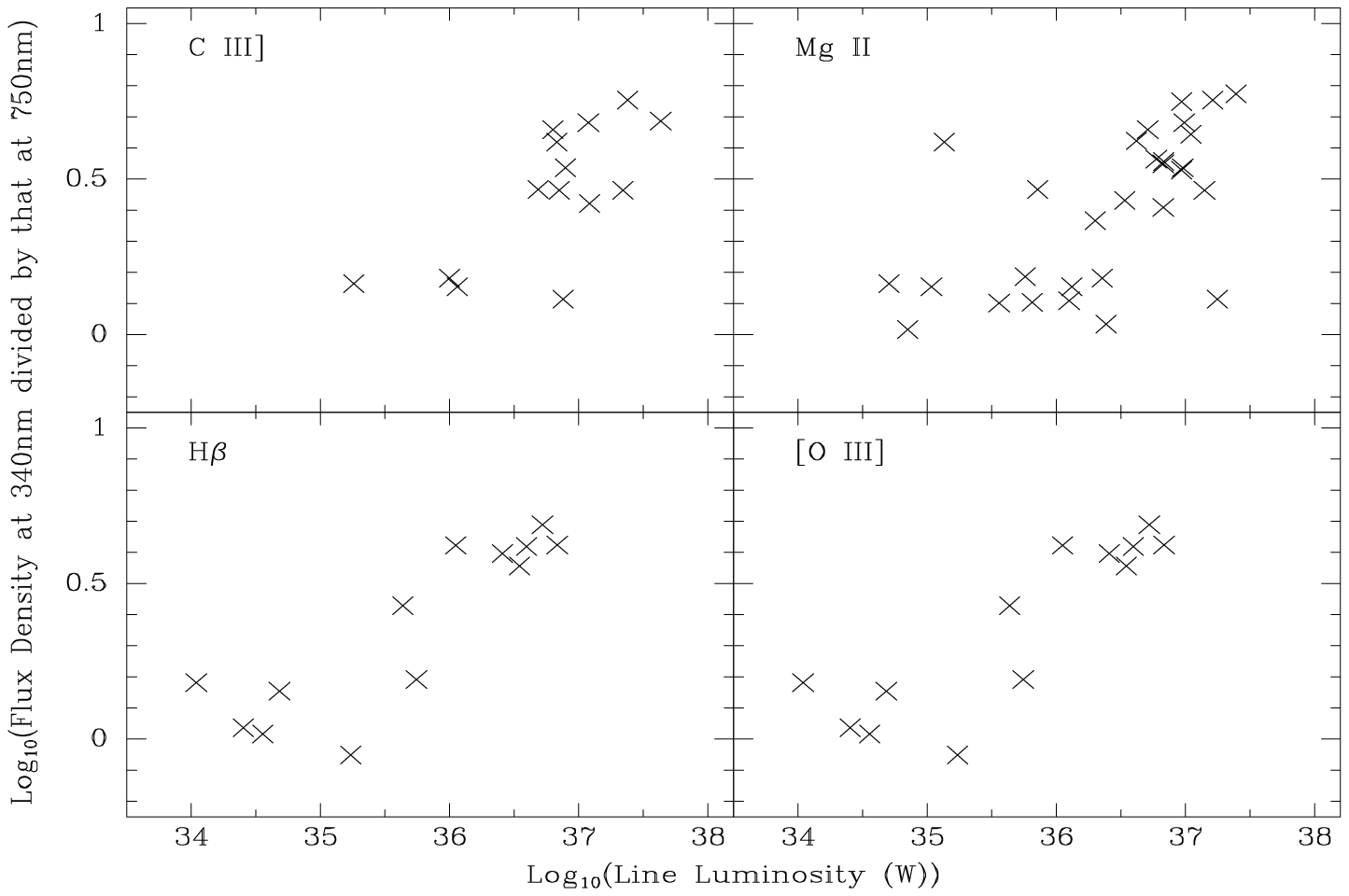,width=180mm}
 \caption{The continuum slope, measured as a function of the 
luminosity of four emission lines. The continuum slope is measured by the
flux ($F_{\lambda}$) at rest-frame 340 nm divided by that at 750nm.
\label{slope_linelum}}

\end{figure}

A stronger correlation is seen between the {\em  emission-line} luminosity 
and the continuum slope (Fig~\ref{slope_linelum}).
The correlation with C~III] luminosity is only significant at the 97\% level,
but the other correlations shown are significant with 99.99\% confidence or
better.

\section{Models\label{models}}

\subsection{Weak Blue Bump}

How should the strength of the Big Blue Bump affect the emission-line
properties? The Big Blue Bump is normally ascribed to thermal emission from 
an accretion disk. The disk could be different in red and blue quasars.
Alternatively, the rest-frame UV/optical emission may be entirely 
synchrotron. Red and blue quasars could have different electron energy
spectra in the jet.
Either way, the emission-lines of red and blue quasars
should be quite different, as the Big Blue Bump emission dominates their  
photoionisation.

We modelled the effect of varying the Big Blue Bump size on the emission
line, using the Cloudy photoionisation code (Ferland 1996). The broad
emission-line region was modelled as a single plane-parallel slab of uniform
hydrogen number density $3.1 \times 10^{16}{\rm m}^{-3}$, and solar
metallicity. The flux incident on the front of the slab at 500nm wavelength
was fixed at 
$F_{\nu} = 1.0\times 10^{-9}{\rm W\ Hz}^{-1}$ (low ionisation model),
$F_{\nu} = 6.3\times 10^{-9}{\rm W\ Hz}^{-1}$ (medium ionisation model) and
$F_{\nu} = 4.0\times 10^{-8}{\rm W\ Hz}^{-1}$ (high ionisation model).
All these parameters are typical of published broad-line region models.

Cloudy's standard AGN photoionising continuum was used. This is 
a power-law in the UV and optical, with exponential
upper- and lower cut-offs, together with a second power-law in the X-rays:
\[
f_{\nu} = \nu^{\alpha_{\rm UV}}e^{\frac{-h\nu}{kT_{\rm BB}}}
e^{\frac{-kT_{\rm IR}}{h\nu}} +a \nu^{\alpha_{\rm X}}
\]
where $T_{\rm BB} = 150,000$K, $kT_{\rm IR} = 0.01$ Ryd and 
$\alpha_{\rm X} = -1$.
The constant $a$ is set such that the slope of a power-law extrapolated 
between 2 keV and 250nm rest-frame, $\alpha_{OX}$, is $-1.4$. Thus the
hard X-ray fluxes of red and blue quasars are same, as suggested by the
ROSAT observations of Siebert et al. (1998).

\begin{figure}
 \psfig{file=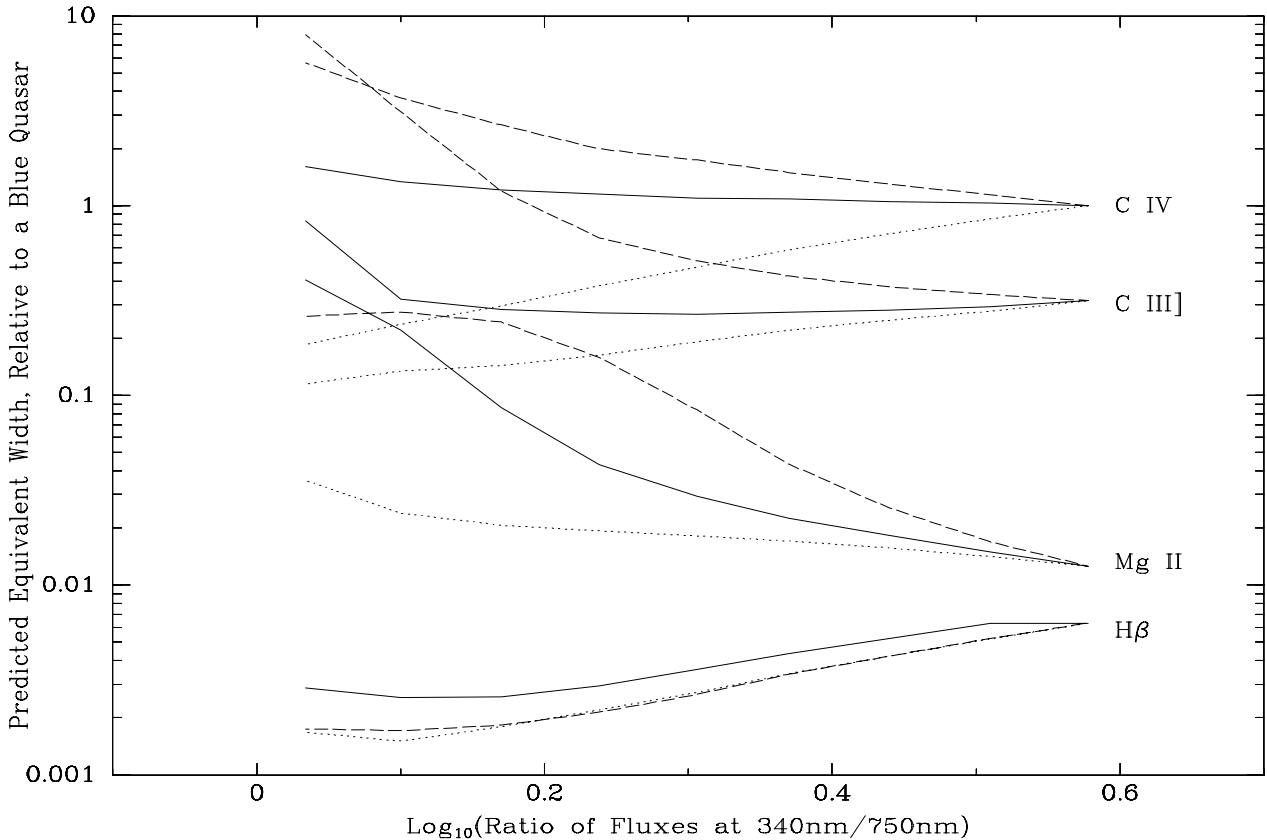,width=180mm}
 \caption{The predicted equivalent widths of C~IV, C~III], Mg~II and
H$\beta$, as a function of Big Blue Bump shape. Big Blue Bump shape is
parameterised by the rest-frame UV/optical continuum slope. The equivalent 
widths are normalised to those
of the bluest quasar models, and a vertical offset has been applied to the 
different lines to improve legibility. The vertical scale is thus arbitrary.
Dotted lines are for the low
ionisation model, solid lines for the medium ionisation model, and dashed lines
for the high ionisation model.
\label{BBB}}

\end{figure}

All that we varied was the strength of the Big Blue Bump, parameterised
by $\alpha_{\rm UV}$. The equivalent widths of the strongest emission lines
were calculated as a function of $\alpha_{\rm UV}$, for each of the three
ionisation models. Results are shown in Fig~\ref{BBB}.

A few clear patterns emerge. H$\beta$ is weaker in weak bump quasars. This
is expected, as it is predominantly photoionised by flux just shortward of
the Lyman limit, which is dominated by the Big Blue Bump. C~III] and C~IV can 
be either weaker or stronger in weak bump quasars, depending on the 
ionisation state of the broad-line region. This too is expected, as the
strength of the Big Blue Bump affects the ionisation parameter, and hence the
population of the different ionisation states of carbon. Mg~II becomes
much stronger in weak bump quasars. Mg~II is a relatively low ionisation
line, and is predominantly excited by X-rays penetrating deep into the
emission-line clouds. Thus an ionising continuum weak in UV photons but
relatively strong in harder X-rays is ideal for Mg~II formation.

A more realistic model would allow for the stratified nature of the 
emission-line regions. It is also unlikely that the physical 
properties of the emission line region are uncorrelated with the colour
of the continuum emission.

\subsection{Dust}

Normal dust models predict a strongly curved continuum in red quasars,
and hence are not consistent with the photometry of the pseudo power-law
sources (FWW). If the dust were deficient in small grains, however, the
UV absorption would be reduced, and the predicted continuum emission could be 
brought into agreement with that observed.

Dust should affect the emission-line ratios, but should not affect the
emission-line equivalent widths, as it should absorb line and continuum
radiation equally. The exception to this would be where the dust is either
patchy on scales comparable to the broad-line region, or where it is mixed in
with the broad-line region.

\subsection{Synchrotron Emission} 

Unified models suggest that
we are observing flat radio spectrum quasars from close to the axis of
their jets. Any emission from these jets would thus be 
relativistically boosted. If these jets sometimes emitted a synchrotron 
component with very red colours in the optical, it could produce 
the red colours. Whiting, Webster \& Francis (2001, hereafter WWF) 
showed that a model of the continuum emission with two components
(a big blue bump and a red synchrotron component) can reproduce the
colours of the pseudo power-law sources as measured by FWW, including the
small (but significant) deviations from a pure power-law. The model also
correctly predicts the polarisation properties of the small fraction of
PHFS quasars for which this data is available.

If the synchrotron model is correct, there
should be an anti-correlation between the equivalent widths of the
emission lines and the redness of the continuum emission. This is
because the emission-line flux of a quasar should be proportional to the 
strength of the Big Blue Bump that photoionises it, and should not be affected
by the strength of any red synchrotron emission component. The red component 
will have too little flux in the UV to ionise emission lines,
and it is probably beamed, and hence will only
illuminate a small part of the emission-line region. Thus if the observed 
red continuum emission greatly exceeds the big blue bump, it will also
greatly exceed the fluxes of the emission lines, which will hence have
very low equivalent widths. The effect should be most marked for longer
wavelength lines, as the red synchrotron component will be relatively stronger
at longer wavelengths. Line ratios should not correlate with
colour.

\section{Discussion\label{discussion}}

At short wavelengths, there are no significant differences between
the emission-line equivalent widths of the red and blue sub-samples.
H$\beta$, however, shows a very significant difference, in the sense that 
most red sources have very weak lines. The data from [O~III] are also
consistent with this (Table~\ref{medians}).

\begin{table}
\begin{center}
\caption{Median Equivalent Widths \label{medians}}
\begin{tabular}{lccc}
\hline 
Line & Blue PHFS Quasars & Red PHFS Quasars & Ratio Blue/Red \\
\hline \\
C~IV         & 3.3 nm & 2.7 nm & 1.2         \\
C~III]       & 2.2 nm & 1.1 nm & 2.0         \\
Mg~II        & 2.6 nm & 2.3 nm & 1.2         \\
H$\beta$     & 7.9 nm & 0.9 nm & 9.1         \\
$[$O~III$]$  & 4.4 nm & 0.4 nm & 12.6        \\
\hline \\
\end{tabular}
\end{center}
\end{table}

These results are consistent with the red synchrotron component
model, if this red component is about ten times stronger than the blue
component at $\sim 500$nm wavelength, but no longer swamps the blue component 
at 300nm. This implies a spectrum at least as red as a power-law of the form 
$F_{\nu} \propto \nu^{\sim -2.8}$, which is consistent with the theoretical
modelling of WWF.

These results are not consistent with the dust model, which predicts no
correlation between equivalent width and colour. Neither are they consistent
with the weak Blue Bump model, which predicts that Mg~II should be
stronger in the red quasars, and that H$\beta$ should only be weaker by 
a factor of $\sim 3$, not $\sim 10$.

Let us therefore assume that the weakness of the lines of the red sub-sample is
due primarily to dilution by a red synchrotron component. 
We detect the H$\beta$ line at $3 \sigma$ confidence or better in all but 
one of the red quasars with decent data. The equivalent widths span roughly 
two orders of magnitude at rest frame 500nm, with no obvious bimodality. 
In at most 10\% of the red quasars can the red continuum component
exceed the blue component by a factor of more than 100, at rest-frame 500nm. 
This too is consistent with the modelling of WWF.

\subsection{Red, Strong-Lined Quasars}

\begin{figure}
\psfig{file=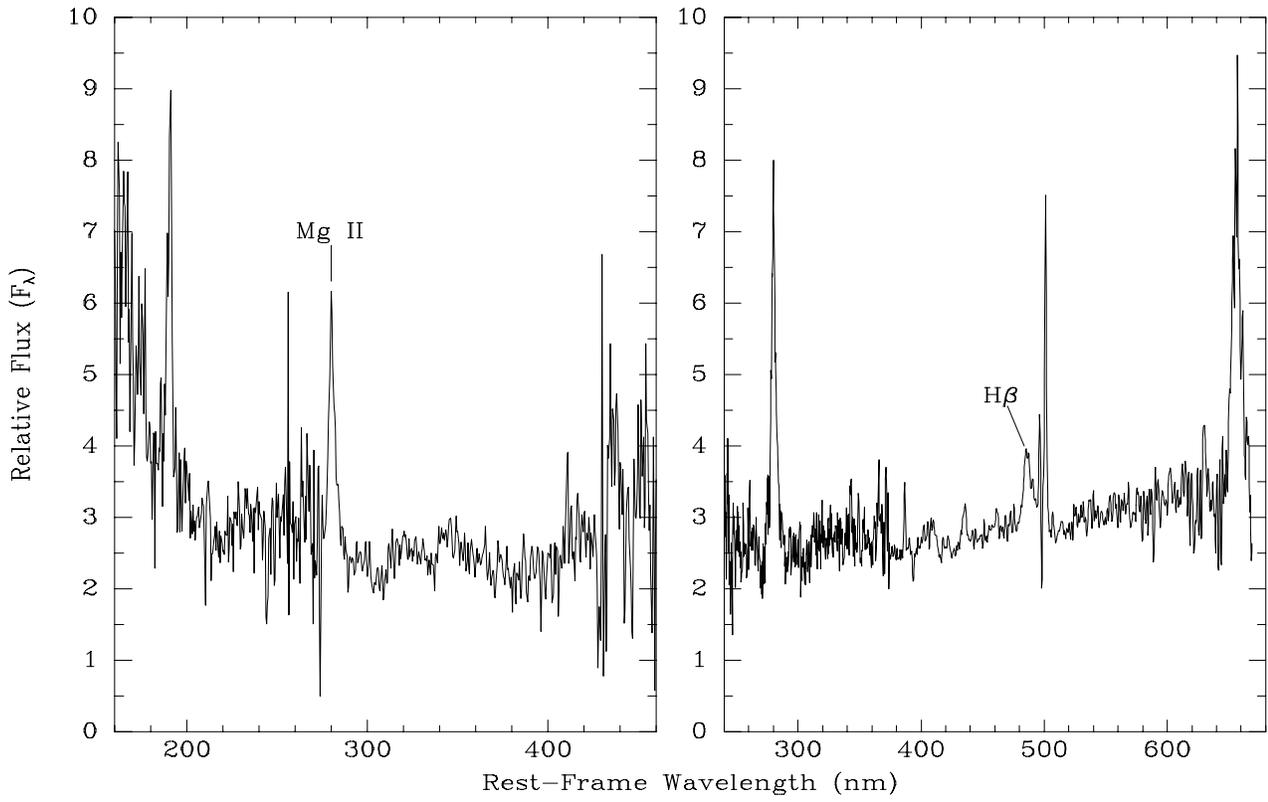,width=180mm}
\caption{Spectra of the quasars from the red sub-sample with the
highest equivalent width H$\beta$ (left) and Mg~II (right) lines.
\label{strongred}}

\end{figure}

If the synchrotron model is correct, all red quasars should have low
equivalent width emission-lines, and all quasars with very low
equivalent widths should be red. The latter is consistent with our
data, but as Fig~\ref{ewbk} shows, there
are a small number of extremely red quasars with high equivalent width
Mg~II lines.
The spectra of the two most extreme red, strong-lined quasars are
shown in Fig~\ref{strongred}. Indeed, these red quasars have the highest
equivalent width Mg~II emission-lines in the sample.

These sources clearly do not fit into the simple synchrotron model.
Dust could redden the continuum without changing the equivalent widths. The 
Balmer decrements of these sources are, however, $\sim 3.5$, which is 
typical of unreddened AGN.

A weak Big Blue Bump model fits the red colours and high Mg~II
equivalent widths of these sources well. We therefore hypothesise that
these quasars have intrinsically weak Big Blue Bumps. Note, however, that
while the equivalent widths of these sources are anomalous, their
line luminosities fit on the correlation with continuum slope
(Fig~\ref{slope_linelum}). 

\subsection{The Correlation between Luminosity and Continuum Slope}

The synchrotron model appears to best fit most of the spectra. What then
determines the relative strength of the synchrotron and Big Blue Bump
components, and hence the quasar colours?
The continuum slope measured between 340 and 750nm should be a good measure
of the relative strength of the synchrotron and Big Blue Bump components.
Emission-line luminosities should correlate strongly with the luminosity
of the Big Blue Bump which photoionises them. Thus
the correlations shown in both Figs~\ref{slope_lum} and \ref{slope_linelum}
suggest that the ratio of synchrotron emission to Big Blue Bump flux is
relatively high in quasars with low luminosity Big Blue Bumps.

This implies that in quasars with low luminosity Big Blue Bumps, the 
fraction
of the accretion energy going into the jet is relatively high, or that the
jet emission extends to shorter wavelengths in these sources. The $1 \sigma$
scatter in this correlation is 0.4 dex, which is better than the scatter
in the Baldwin Effect. This correlation may thus be useful for cosmological
distance scale work.

\subsection{Implications for BL Lac Objects}

BL Lac objects are normally defined as having no observed emission lines
with rest-frame equivalent widths greater than 0.5nm (eg. Stickel et al. 1991,
Rector et al. 2000). In our model, this implies that the red synchrotron 
emission component exceeds the blue one at the line wavelength by a factor
of $\sim 10$.

We do not see any bimodality in the distribution of equivalent factors, though
our sub-sample is small. If confirmed, this would imply that radio
selected BL Lac objects are simply the tail of the normal population of
flat radio spectrum quasars, and not a different class with intrinsically
weaker emission lines.

Our analysis suggests that the synchrotron emission component is very red,
and hence that it dilutes long wavelength lines far more effectively than
short wavelength lines. This implies that BL Lac samples are 
biassed against finding high redshift objects. 

How significant could this bias be? Consider a model in which the red 
continuum
component has a spectrum of the form $F_{\nu} \propto \nu^{-2.8}$, and 
the logarithm of the ratio of red to blue components at the wavelength of 
H$\beta$ is uniformly distributed between 0 and 2, as suggested by our 
observations. The ratio of red 
to blue components at the wavelength of Mg~II will be 3.4 times smaller than
at H$\beta$, while at C~III] it will be 7.8 times smaller. It is assumed
that the ratio of red to blue components does not correlate with redshift
or luminosity.

Now consider, for example, a sample of  
BL Lac objects chosen on the basis of spectra covering 400 ---
600nm. At redshift zero, the strongest line in this wavelength range
will be H$\beta$, which in blue PHFS quasars has a rest-frame equivalent
width of $\sim 5$nm. For this line to have an observed equivalent
width of less than 0.5nm, we require that the red component at the
wavelength of H$\beta$ exceed the blue component by a factor of ten. According
to our model, 50\% of all red 
quasars meet this criterion and would hence be 
classed as BL Lac objects.

At redshift one, however, Mg~II will be the strongest line within this
observed wavelength region. In blue PHFS quasars, Mg~II has a median
rest-frame equivalent width of 3nm. For the line to have a rest-frame 
equivalent width of less than 0.5nm, we require that the red continuum
component exceed the blue one by a factor of 6 at the wavelength
of Mg~II. This implies that the red component must exceed the blue by
a factor of $3.4 \times 6 = 20$ at the wavelength of H$\beta$. According to
our model, only 35\% of red quasars will meet this criterion. Repeating this 
calculation at redshift two (where C~IV is 
the strongest line), only 6\% of red PHFS quasars would be classed as 
BL Lac objects.

Thus any sample using spectra at a uniform observed-frame wavelength
will miss $\sim 30$\% of the BL Lacs at redshift one, and $\sim 90$\%
at redshift two. Our small sub-sample size makes these estimates
highly uncertain. Nonetheless, this is probably sufficient to cancel out 
the negative evolution
claimed by some authors (eg. Rector et al. 2000), but not to match the
enormous positive evolution seen in other classes of quasar.

\subsection{Differences between Radio-Loud and Radio-Quiet Quasars}

It has long been known that the optical spectra of radio-loud
and radio-quiet quasars are remarkably similar, despite the
enormous difference in their radio luminosities. A number of
small differences in the emission-line properties have been claimed
(eg. Boroson \& Green 1992, Francis, Hooper \& Impey 1993, Corbin \& 
Francis 1994, Corbin 1997), but only on the basis of small, poorly matched 
and/or incomplete samples. These studies 
found that C~IV, C~III] and [O~III] had higher equivalent widths in flat 
radio spectrum quasars than in radio-quiet quasars.

The optical luminosity and redshift distribution of the LBQS are
quite similar to those of the PHFS quasars, so any differences
between the emission-lines of the two samples must be physical.
If we compare the spectra of the LBQS QSOs with the blue PHFS sub-sample, we
find that the equivalent width distributions of C~IV, H$\beta$ and
[O~III] are significantly different, with 95\% confidence. C~III]
is different with 94.8\% confidence. In all cases, the equivalent 
widths of the blue PHFS quasars are greater than those of the LBQS QSOs.
The red PHFS quasars have significantly weaker Mg~II and H$\beta$ 
equivalent widths than the LBQS QSOs, presumably due to dilution by the
red synchrotron component.

\section{Conclusions\label{conclusions}}

We conclude that the spectra of the red PHFS quasars are significantly
different from those of the blue PHFS quasars. Small number statistics
make our conclusions tentative, though they  are formally significant. 
The probable differences are
consistent with a model in which the red colours are due to the addition
of a featureless red synchrotron continuum component to an otherwise normal
blue quasar spectrum, as proposed by WWF. The red component must have
a spectrum at least as red as a power-law of the form 
$F_{\nu} \propto \nu^{-2.8}$. This
red component contributes no more than half the continuum flux at
rest-frame 300nm, but at rest-frame 500nm it contributes about
90\% of the continuum flux of the red quasars. The physics of such a 
component is discussed by WWF.

The relative strengths of the blue and red components span two orders of
magnitude at rest-frame 500nm. The blue component is relatively weaker in
low optical luminosity sources.

If this model is correct, then existing BL Lac surveys are biassed against
high redshift objects, though this bias is insufficient to bring the
evolution of BL Lacs into consistency with the evolution of other quasars.

A few PHFS quasars, however, have both very red colours and very high
equivalent width emission-lines. These quasars do not easily fit the 
synchrotron model, but their properties can be fit by an intrinsically
weak Big Blue Bump model.
We confirm that the emission-lines of optically selected QSOs have
significantly weaker equivalent widths than those of radio-loud,
flat radio spectrum quasars with similar optical luminosities.

\bigskip

Digital copies of the composite spectra and of the measured equivalent widths
are available from the on-line copy of this paper.

\section*{Acknowledgements}

We wish to thank Belinda Wilkes for making her spectra available to
us in digital format, and Chriss Wallwork for doing a literature
search for new redshifts.

\section*{References}

\reference Boroson, T. \& Green, R. 1992, ApJ, 338, 630

\reference Corbin, M.R. 1997, ApJS, 113, 245

\reference Corbin, M.R. \& Francis, P.J. 1994, AJ, 108, 2016

\reference Drinkwater, M.J., Webster, R.L., Francis, P.J., Condon, J.J.,
Ellison, S.L., Jauncey, D.L., Lovell, J., Peterson, B.A.,
\& Savage, A. 1997, MNRAS, 284, 85 

\reference Falomo, R. 1991, AJ, 102, 1991

\reference Ferland, G.J. 1996, Hazy, a brief introduction to Cloudy,
University of Kentucky Department of Physics and Astronomy internal
report.

\reference Francis, P.J., Hooper, E.J. \& Impey, C.D. 1993, AJ, 106, 417

\reference Francis, P.J., Whiting, M.T. \& Webster, R.L. 2000,
PASA 17, 56 (FWW)

\reference Malkan, M.A. \& Sargent, W.L.W. 1982, ApJ, 254, 22

\reference Masci, F.J., Webster, R.L. \& Francis, P.J. 1998,
MNRAS 301, 975

\reference McDowell, J.C., Elvis, M., Wilkes, B.J., Willner, S.P.,
Oey, M.S., Polomski, E., Bechtold, J. \& Green, R.F. 1989, ApJL, 345, L13

\reference Morris S.L., Weymann R.J., Anderson S.F., Hewett P.C., 
Foltz C.B., Chaffee F.H. \&
Francis P.J. 1991, AJ, 102, 1627

\reference Rector, J.A., Stocke, J.T., Perlman, E.S., Morris, S.L. \&
Gioia, I.M. 2000, AJ 120, 1626

\reference Rieke, G.H., Lebofsky, M.J. \& Wisniewskiy, W.A. 1982, ApJ, 
263, 73

\reference Serjeant, S. \& Rawlings, S. 1997, Nature, 379, 304

\reference Siebert, J., Brinkmann, W., Drinkwater, M.J., Yuan, W.,
Francis, P.J., Peterson, B.A. \& Webster, R.L. 1998, MNRAS, 301, 261

\reference Stickel, M., Padovani, P., Urry, C.M., Fried, J.W.,
K\"uhr, H. 1991, ApJ, 374, 431

\reference Webster, R.L., Francis, P.J., Peterson, B.A.,
Drinkwater, M.J., \& Masci, F.J. 1995, Nature, 375, 469

\reference Whiting, M.T., Webster, R.L. \& Francis, P.J. 2000, MNRAS
in press (WWF)

\reference Wilkes, B.J., Wright, A.E., Jauncey, D.L. \&
Peterson, B.A., 1983, PASA, 5, 2

\end{document}